\documentclass[prl,twocolumn,floatfix,showpacs]{revtex4-1}

\usepackage{graphicx}
\usepackage[latin1]{inputenc}
\usepackage{dcolumn}
\usepackage{bm}

\begin{document}

\title{Current dependence of the huge negative magnetoresistance in high-mobility two-dimensional electron gases}

\author{L. Bockhorn$^1$}
\email{bockhorn@nano.uni-hannover.de}
\author{J. I\~narrea$^2$}
\author{R. J. Haug$^1$}

\affiliation{$^1$Institut f\"ur Festk\"orperphysik, Leibniz Universit\"at Hannover, Hannover, Germany\\
$^2$Escuela Polit$\acute e$cnica Superior, Universidad Carlos III de Madrid, Madrid, Spain}

\date{\today}

\begin{abstract}
In high-mobility two-dimensional electron gases Landau levels are already formed at very small magnetic field values. Such two-dimensional electron gases show a huge negative magnetoresistance at low temperatures and an unexpected and very strong non-linear behavior with the applied current. This non-linearity depends on carrier concentration and is explained by the subtle interplay of elastic scattering within Landau levels and in between Landau levels.

\end{abstract}


\maketitle

A negative magnetoresistance with parabolic magnetic field dependence at low temperatures was firstly observed by Paalanen~\textit{et~al.}~\cite{Paalanen1983} for a two dimensional electron gas (2DEG). This negative magnetoresistance showed temperature dependence and was theoretical described by the electron-electron interaction correction to the conductivity~\cite{Houghton1982, Girvin1982}. Since the first observation of this effect the sample quality and also the mobility of 2DEG improved quite drastically. This has allowed to observe a more pronounced negative magnetoresistance~\cite{Dai2010, Bockhorn2011, Hatke2012, Mani2013, Bockhorn2013, Bockhorn2013a, Bockhorn2014} which is often called huge magnetoresistance. In addition to this huge magnetoresistance a peak around zero magnetic field is also examined in high-mobility 2DEG. Recently, it was shown that this peak originates from an interplay of smooth disorder and macroscopic defects~\cite{Mirlin2001, Polyakov2001, Bockhorn2014}. In contrast, the huge magnetoresistance is not fully understood so far for high-mobility 2DEGs. The behavior of the huge magnetoresistance is effected by different scattering events, e.~g. interface roughness, rare strong scatterers, background impurities and remote ionized impurities, which makes a theoretically description more complex. The main scattering event is given by remote ionized impurities due to the fact that the crossover between the huge magnetoresistance and the peak is dominated by this type of disorder~\cite{Bockhorn2014}.

The behavior of the huge magnetoresistance which is characterized by its curvature was already analyzed for different conditions. At first a strong temperature dependence was detected at low temperatures ($T<1\,$K)~\cite{Bockhorn2011, Bockhorn2013a}. An accurate analysis of the curvature shows two different temperature dependences: For strong magnetic fields ($\omega_c\,\tau>$1) the value of the curvature decreases with temperature by $T^{-1/2}$ and for weak magnetic fields ($\omega_c\,\tau<$1) the temperature dependence is given by $T^{-1}$~\cite{Bockhorn2013a}. Furthermore, the huge magnetoresistance gets more pronounced by decreasing the electron density while its curvature is left unchanged~\cite{Bockhorn2011}. Additionally, effects of an in-plane magnetic field component were examined~\cite{Hatke2012, Bockhorn2014}. Based on these results the electron-electron interaction correction to the conductivity considering mixed disorder~\cite{Gornyi2003, Gornyi2004} was assumed as an explanation for the huge magnetoresistance~\cite{Li2003}. However, different groups previously reported on discrepancies between such theoretical models and the observed huge magnetoresistance~\cite{Bockhorn2011, Hatke2012, Mani2013}. 

\begin{figure}
   \centering
   \includegraphics{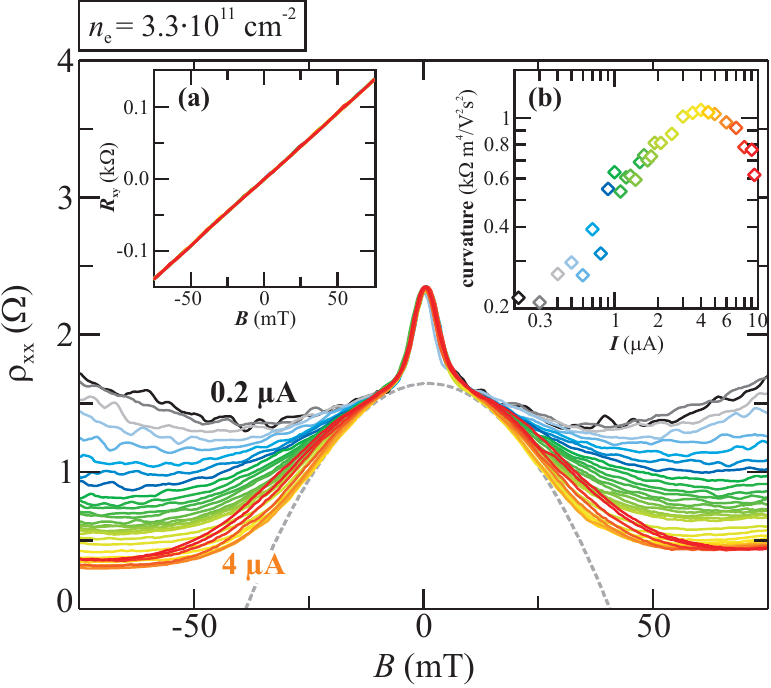}
   \caption{\label{fig:High}The longitudinal resistivity$\,\rho_{xx}$ is plotted vs. magnetic field$\,B$ for different currents$\,I$ ranging from 0.2$\,\mu$A to 10$\,\mu$A at $T=100\,$mK. The huge magnetoresistance (marked by a dashed grey parabola) gets more pronounced by increasing the current$\,I$ while the peak around zero magnetic field is left unchanged. Also, the Hall resistance~$R_{xy}$ is unaffected by different currents as seen in the inset (a). The inset (b) shows the corresponding curvature of the huge magnetoresistance vs. current$\,I$ on a log-log scale.}
\end{figure}

In the following we present measurements of the huge magnetoresistance as a function of applied current~$I$ and introduce a theoretical model which was originally developed for the description of  microwave-induced resistance oscillations (MIRO)~\cite{Zudov2001, Ye2001} respectively for a phenomenologically similar effect, the Hall field-induced resistance oscillations (HIRO) ~\cite{Yang2002}. Both MIRO and HIRO are explained by elastic scattering between high Landau levels due to remote ionized impurities~\cite{Durst2003}. We show that simulations of this theoretical model also describe the behavior of the huge magnetoresistance qualitatively correct.

We studied high-mobility 2DEGs which were realized in 30$\,$nm wide GaAs~quantum wells grown by molecular-beam epitaxy. The quantum wells are cladded by an undoped \mbox{Al$_{0.25}\,$Ga$_{0.75}\,$As}~spacer and a \mbox{GaAs/Al$_{0.25}\,$Ga$_{0.75}\,$As}~superlattice. Additionally, they are Si$\:\delta$-doped from both sides. The highest electron mobility of \mbox{$\mu\approx11.9\cdot10^{6}\,$cm$^{2}$/Vs} was measured for \mbox{$n_{e}\approx3.3\cdot10^{11}\,$cm$^{-2}$}. Details on our high-mobility 2DEGs are also found in Ref.~\cite{Bockhorn2011, Bockhorn2013, Bockhorn2013a, Bockhorn2014}. The specimens were structured as Hall bars with a total length of 1.8$\,$mm, a width of $w=0.2\,$mm and a potential probe spacing of $l=0.3\,$mm. The Hall bars were defined by photolithography and wet etching. We performed all magnetotransport measurements in a dilution refrigerator with a base temperature of 20$\,$mK by using low-frequency (13$\,$Hz) lock-in technique. The applied current$\,I$ oscillates with the same frequency and was varied by changing amplitude. Various ungated and gated samples were used for our magnetotransport measurements. The behavior of the negative magnetoresistance was always similar.

To examine the behavior of the negative magnetoresistance for different electric fields the applied current$\,I$ was varied for a given temperature$\,T$ and an given electron density$\,n_e$. These current dependent measurements were repeated for various electron densities. Figure~\ref{fig:High} shows the longitudinal resistivity$\,\rho_{xx}$ vs. magnetic field$\,B$ for different currents$\,I$  ranging between 0.2$\,\mu$A and 10$\,\mu$A for \mbox{$n_e=3.3\cdot10^{11}\,$cm$^{-2}$}. The strong negative magnetoresistance clearly consists of two contributions which differ in their dependence on the applied current. The huge magnetoresistance (dashed grey parabola) gets more pronounced by increasing the current while the peak around zero magnetic field is left unchanged. At the same time, the Hall resistance~$R_{xy}$ is left unchanged for different currents as seen in the inset~(a) of Fig.~\ref{fig:High}. The inset~(b) illustrates a detailed analysis of this curvature for several currents. Here, the curvature of a parabola fitted to the experimental data is plotted vs. current~$I$. Firstly, the curvature value clearly increases with current for $\,I\le\,4\,\mu$A and accordingly the huge magnetoresistance gets more pronounced. The strongest development of the huge magnetoresistance is observed around $I=4\,\mu$A. Correspondently, a maximum is detected for the curvature value. However, the huge magnetoresistance vanishes for further increase of the current$\,I\ge\,4\,\mu$A and the corresponding curvature value decreases.

It was previously shown that the huge magnetoresistance rapidly vanishes with temperature and the corresponding curvature value decreases~\cite{Bockhorn2011, Hatke2012}. Therefore, the current dependence of the huge magnetoresistance observed up to $I=4\,\mu$A is clearly not caused by a heating effect. We determined the electron temperature from the Shubnikov-de~Haas oscillations~(SdHO) to prove this assumption following  \mbox{Coleridge~\textit{et~al.}~\cite{Coleridge1989}}. Despite the strong current dependence of the huge magnetoresistance we found a constant electron temperature of $T\sim100\,$mK.

\begin{figure}
   \centering
   \includegraphics{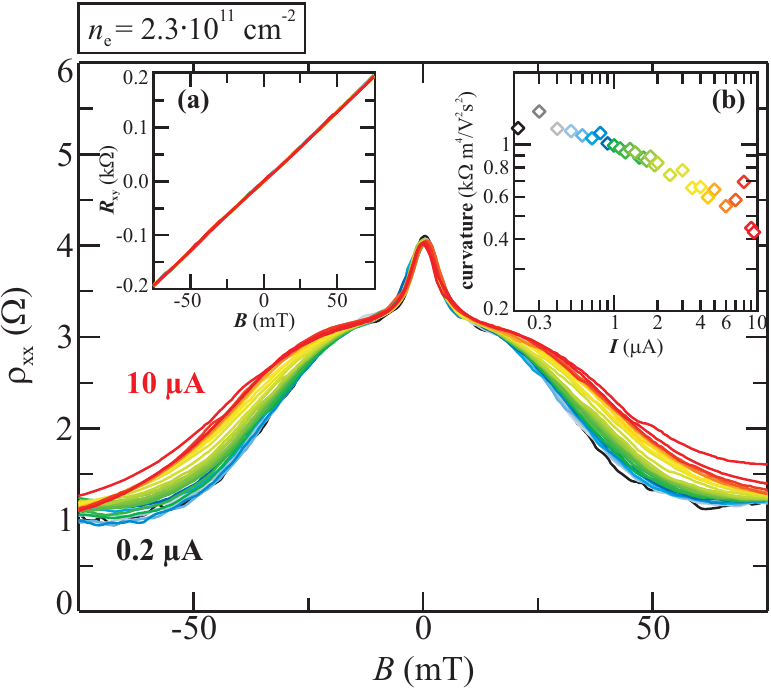}
   \caption{\label{fig:Low}The longitudinal resistivity is shown$\,\rho_{xx}$ vs. magnetic field$\,B$ for different currents ranging from 0.2$\,\mu$A to 10$\,\mu$A at $T=100\,$mK. The huge magnetoresistance slightly decreases with current$\,I$. The inset (a) shows the Hall resistance~$R_{xy}$ vs. magnetic field~B for several currents. The corresponding curvature of the huge magnetoresistance is plotted vs. current$\,I$ on a log-log scale in inset (b). The curvature value clearly decreases with current.}
\end{figure}

Further current dependent measurements for several electron densities showed that the nonlinear behavior of the huge magnetoresistance is related to the electron density$\,n_e$. Figure~\ref{fig:High} reveals that for \mbox{$n_e=3.3\cdot10^{11}\,$cm$^{-2}$} the huge magnetoresistance gets more pronounced by increasing current$\,I$. In contrast, a different dependence is observed for \mbox{$n_e=2.3\cdot10^{11}\,$cm$^{-2}$}. The longitudinal resistivity$\,\rho_{xx}$ is plotted vs. magnetic field$\,B$ in Fig.~\ref{fig:Low}. Firstly, the larger height of the huge magnetoresistance is noticed. This could be explained by the lower electron density due to the fact that the huge magnetoresistance gets more pronounced by decreasing the electron density~\cite{Bockhorn2011}. Here, the Hall resistance~$R_{xy}$ is also unaffected by different currents as seen in inset~(a). However, for \mbox{$n_e=2.3\cdot10^{11}\,$cm$^{-2}$} the huge magnetoresistance vanishes slightly by increasing the current while the peak around zero magnetic field is still left unchanged. This observation is also seen in the dependence of the curvature. Inset~(b) shows that the curvature value decreases with current$\,I$. In this situation, the current dependence of the huge magnetoresistance is different for lower electron densities than for $n_e=3.3\cdot10^{11}\,$cm$^{-2}$.

Measurements considering nonlinear transport mechanisms at non quantizing magnetic fields were recently presented by Shi~\textit{et~al.}~\cite{Shi2014}. They analyzed electric field dependences of a more pronounced negative magnetoresistance around zero magnetic field up to $|B|\sim100\,$mT for similar mobilities. However, their negative magnetoresistance shows no parabolic magnetic field dependence and persists to higher temperatures than the huge magnetoresistance observed in our high-mobility 2DEGs. Also, Zhang~\textit{et~al.}~\cite{Zhang2009} reported on nonlinear behavior of the longitudinal resistivity for different electric fields. On basis of the much lower mobilities they assumed inelastic scattering in overlapping Landau levels at small magnetic fields. In contrast, in our high-mobility 2DEGs the Landau levels are separated in the considered magnetic field range. Our observed dependence of the huge magnetoresistance on the electric field deviates from common theories (see e.~g.~\cite{Dmitriev2012}) studying a negative magnetoresistance.

We consider a recently introduced theoretical model~\cite{Inarrea2014} to describe the huge magnetoresistance for varying currents due to the fact that the theoretically expected curvature values fit to our experimental data for several conditions (e.~g. temperature, in-plane magnetic field component). Here, a former transport model considering MIRO and HIRO (see e.~g.~\cite{Dmitriev2012}) is adapted to the situation of the huge magnetoresistance which is based on elastic scattering between Landau levels due to remote ionized impurities~\cite{ridley,ina1, ando}. Considering references~\cite{Inarrea2014,ina1, ridley} the longitudinal resistivity$\,\rho_{xx}$ is then expressed by
\begin{equation}
\rho_{xx}\propto\Bigg\{ \frac{1-e^{\left[-\frac{2\pi\Gamma }{\hbar w_{c}}\right]} } {1-2e^{\left[-\frac{\pi\Gamma }{\hbar w_{c}}\right]}   \cos \left[\frac{2\pi  \Delta}{\hbar w_{c}}\right]+e^{\left[-\frac{2\pi\Gamma }{\hbar w_{c}} \right]} } \Bigg\}
\label{eq:1}
\end{equation}
with $\Delta$ being the energy drop along the scattering jump from the initial Landau level to the final Landau level due to a static electric field~$\xi_{DC}$.
 The static electric field$\,\xi_{DC}$ is aligned in current direction and is proportional to the energy drop~$\Delta$ through \mbox{$\Delta= e\xi_{DC}\Delta X^{0}$}. $\Delta X^{0}$ is the average effective distance of electrons for every scattering jump. Thus, the longitudinal resistivity~$\rho_{xx}$ directly depends on the Landau level width~$\Gamma$ as well as on the static electric field~$\xi_{DC}$ which is proportional to the current~$I$. The Landau level width~$\Gamma$ becomes narrow \mbox{($\Gamma\ll\hbar\,\omega_c$)} in high-mobility 2DEGs. According to this the number of states with similar energy in a Landau level rises with decreasing Landau level width$\,\Gamma$, while the density of states between two Landau levels decreases. 

The tilt angle of the Landau levels is important for the description of the huge magnetoresistance for different currents. Considering eq.~(\ref{eq:1}) an increase of static electric field$\,\xi_{DC}$ with the applied current$\,I$ could lead to two opposing effects in terms of the longitudinal resistivity$\,\rho_{xx}$. On the one hand, the tilt of the Landau levels increases with the static electric field$\,\xi_{DC}$. The energy drop$\,\Delta$ rises and scattering events end up between two Landau levels. Here, only a low density of final states is found. The longitudinal resistivity$\,\rho_{xx}$ following eq.~(\ref{eq:1}) becomes smaller which comes along with a more pronounced huge magnetoresistance. On the other hand, a sufficient high current$\,I$ triggers more scattering processes which leads to wider Landau levels. In this case the longitudinal resistivity$\,\rho_{xx}$ considering eq.~(\ref{eq:1}) rises and the huge magnetoresistance starts to vanish.

\begin{figure}
   \centering
   \includegraphics{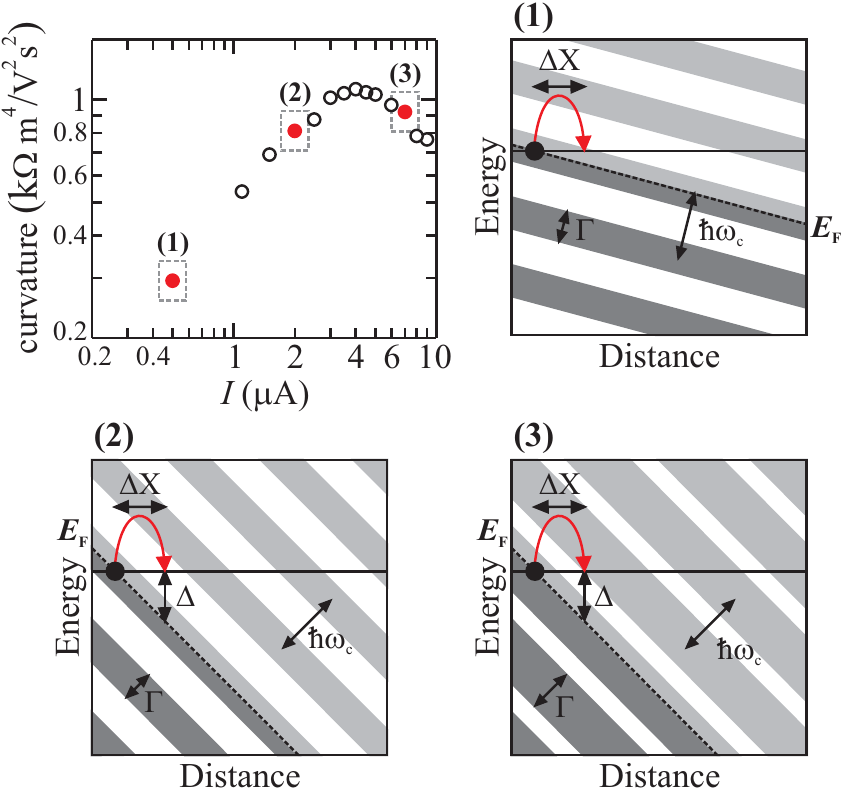}
   \caption{\label{fig:Theorie}The curvature value of the huge magnetoresitance is plotted against the current$\,I$ on a log-log scale for~\mbox{$n_e=3.3\cdot10^{11}\,$cm$^{-2}$}. The circles are experimentally determined values of the curvature. Three different scattering regions are marked. \textbf{(1)}-\textbf{(3)} present the corresponding schematic diagram considering elastic scattering process. The narrow Landau levels are illustrated as grey stripes. \textbf{(1)}~The scattering is close to the center of a Landau level. \textbf{(2)}~The tilt of the Landau levels increases with current and the scattering is shifted to empty states between two Landau levels. \textbf{(3)}~ The increase of the Landau level width becomes important and compensates the tilt.}
\end{figure}

On account of this we are able to explain the elastic scattering process between Landau levels in the situation of various currents for \mbox{$n_e=3.3\cdot10^{11}\,$cm$^{-2}$}. The curvature of the huge magnetoresistance is shown vs. current~$I$ at $T=100\,$mK on a log-log scale in Fig.~\ref{fig:Theorie}. The circles are experimentally determined values of the curvature. Three different scattering regions are marked by numbers and the corresponding schematic diagrams concerning scattering in and between Landau levels are also shown in Fig.~\ref{fig:Theorie}. In region~\textbf{(1)} the current~$I$ is too low for an increase of the Landau level width~$\Gamma$. Hence, the scattering is close to the center of a Landau level. Here, the probability of an empty state is high which corresponds with a large scattering rate and with an increase of the longitudinal resistivity~$\rho_{xx}$. The resulting height of the huge magnetoresistance is small. If the current keeps rising, the tilt of the Landau levels increases and scattering events are shifted to a region with a lower density of states (scattering region~\textbf{(2)}). Now, scattering events might end up between two Landau levels. The scattering rate decreases and the huge magnetoresistance appears which becomes more pronounced by further increase of the current. Thus, for this scattering region the effect of the Landau level tilt is important and dominates the longitudinal resistivity~$\rho_{xx}$. However, around $I=4$ $\mu A$ the scenario changes. The effect of the increasing Landau level width~$\Gamma$ takes over in the scattering region~\textbf{(3)}. The further increased current produces additional scattering events. Hence, the Landau levels become wider and provide a larger number of empty states. Scattering events on several types of disorder become important and are efficient enough to compensate the influence of the Landau level tilt. Consequently, the longitudinal resistivity$\,\rho_{xx}$ rises again and the huge magnetoresistance starts to vanish. Due to the simultaneous action of this two opposing effects the destruction of the huge magnetoresistance is slow. The situation in scattering region~\textbf{(3)} is similar to our observations for lower electron densities~\mbox{$n_e<3.3\cdot10^{11}\,$cm$^{-2}$}.

According to our experimental data we identify two different regimes in the increase of the Landau level width~$\Gamma$ due to the current~$I$. The increase in $\Gamma$ is negligible for sufficiently low currents~$I<4\,\mu$A. However, at $I\sim4\,\mu$A the Landau level width~$\Gamma$ begins to increase. In order to simulate this behavior we have phenomenologically introduced a smoothed step function which relates the increase ($\delta\Gamma$) of the Landau level width with the energy drop~$\Delta$ and the initial Landau level width~$\Gamma_{i}$. Hence, the increase of the Landau level width is expressed by $\delta\Gamma\propto 1+\tanh (\Delta - \Gamma_{i})$. $\Delta<\Gamma_{i}$ and so $\delta\Gamma$ is negligible for low currents. At sufficiently high currents we reach a point where the energy drop~$\Delta$ surpasses the initial Landau level width~$\Gamma_{i}$. $\delta\Gamma$ becomes important and we enter a new scattering regime.

\begin{figure}
   \centering
   \includegraphics{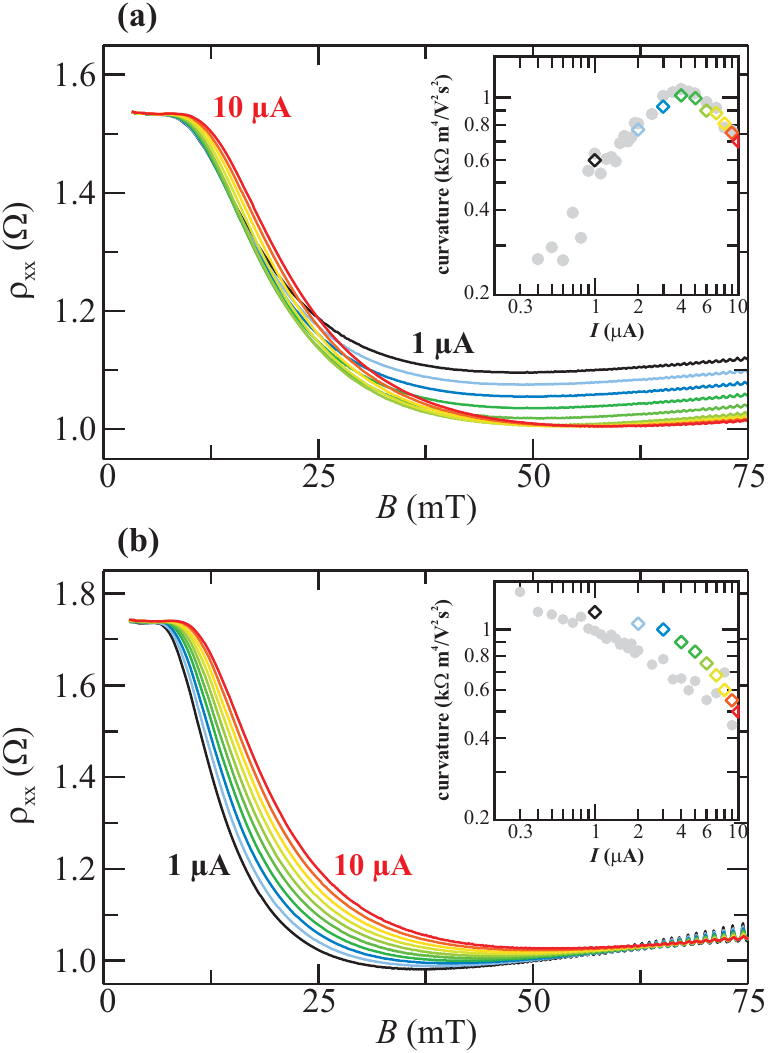}
   \caption{\label{fig:Simulation}The calculated longitudinal resistivity$\,\rho_{xx}$ is plotted against magnetic field$\,B$ for different currents~$I$. Simulations for two different electron densities are presented. Both insets compare the corresponding curvatures of the huge magnetoresistance from these simulations with the experimental data (grey circles) using a log-log scale. (a)~Firstly, the huge magnetoresistance becomes more pronounced with current~$I$ for \mbox{$n_e=3.3\cdot10^{11}\,$cm$^{-2}$}. Here, the simulations fit nicely to our experimental data. (b)~The current dependent behavior of the huge magnetoresistance is shown for \mbox{$n_e=2.3\cdot10^{11}\,$cm$^{-2}$}. The huge magnetoresistance lessens with increasing current which results in a decreasing curvature value.}
\end{figure}

On the basis of the smoothed step function~$\delta\Gamma$ and eq.~(\ref{eq:1}) the longitudinal resistivity~$\rho_{xx}$ was simulated as a function of the current~$I$ for various electron densities~$n_e$. Figure~\ref{fig:Simulation}~(a) presents the simulation of the remarkable behavior of the huge magnetoresistance for \mbox{$n_e =3.3\cdot10^{11}\,$cm$^{-2}$}. The corresponding inset reflects the dependence of the curvature value for a better comparison with the experimental data~(grey circles). In Fig.~\ref{fig:Simulation}~(a) one sees that the huge magnetoresistance gets more pronounced with current~$I$ at the beginning. Respectively, the value of the curvature rises as shown in the inset. At $I=4\,\mu$A a maximum is detected and the huge magnetoresistance decreases for higher currents~$I>4\,\mu$A as expected from the experimental data. Here, the calculated values of the curvature fit nicely to the determined experimental values~(grey circles). Figure~\ref{fig:Simulation}~(b) presents the current dependent behavior of the huge magnetoresistance for \mbox{$n_e =2.3\cdot10^{11}\,$cm$^{-2}$}. The huge magnetoresistance lessens with increasing current~$I$ which results in a decreasing curvature value for this lower electron density (see corresponding inset). The simulation qualitatively describes the observed behavior of the huge magnetoresistance. The observed nonlinear behavior of the huge negative magnetoresistance can be reproduced in simulations by taking into account elastic scattering within Landau levels and in between Landau levels. However, for \mbox{$n_e =2.3\cdot10^{11}\,$cm$^{-2}$} the theoretically expected height of the huge magnetoresistance is lower as the experimental observed height. From this it follows that the huge magnetoresistance depends stronger on the electron density than supposed.

As long as the considered disorder potential is dominated by scatterings events due to remote ionized impurities our theoretical model describes the huge magnetoresistance qualitatively correct. However, the behavior of the huge magnetoresistance is referable to several types of disorder as mentioned above. The probability of scattering on interface roughness, rare strong scatterers, background impurities and remote ionized impurities are in high-mobility samples most likely~\cite{Dmitriev2012}. Recently, it was shown that rare strong scatterers, respectively oval defects, influence the behavior of the longitudinal resistivity at small magnetic fields (see Ref.~\cite{Bockhorn2014} ). Hence, the peak around zero magnetic field reflects the interplay of remote ionized impurities and rare strong scattereres. However, the contribution of rare strong scatterers to the longitudinal resistivity decreases with magnetic field. Already, the crossover between the peak and the huge magnetoresistance is dominated by scatterings events due to remote ionized impurities. The strength of interface roughness could be probed by transport measurements in different crystal directions. An anisotropy between this measurements is a sign for a contribution of interface roughness to the longitudinal resistivity. We observe an anisotropy of about 10$\,\%$ for our high-mobility 2DEGs. Hence, the influence of interface roughness to scattering events cannot be neglect for the description of the huge magnetoresistance. In addition, the contribution of the listed types of disorder rises if the electron density is decreased. According to this also the probability of scattering events on background impurities increases and becomes relevant for the behavior of the huge magnetoresistance. Correspondingly, the huge magnetoresistance is not only dominated by a long range disorder potential due to remote ionized impurities. Here, the interplay of several types of disorder leads to changed scattering mechanisms which is mainly seen in the current dependent behavior of the huge magnetoresistance in the situation of $n_e=2.3\cdot10^{-11}\,$cm$^{-2}$. Until now the interplay of different types of disorder is not satisfactorily considered in common theory models.

In conclusion, we showed that the huge magnetoresistance is influenced by elastic scattering between Landau levels due to remote ionized impurities. On the basis of current dependent measurements we identify two different scattering regimes in the increase of the Landau level width~$\Gamma$. Simulations of the theoretical model fit nicely with the experimental data for higher electron densities. However, we also found that the huge magnetoresistance depends stronger on the interplay of different types of disorder than assumed. So, further calculations are left for future work to explain the strong electron density dependence.

\begin{acknowledgments}
We are grateful to \mbox{D. Schuh}, \mbox{C. Reichl} and \mbox{W. Wegscheider} for providing us with ultra clean high-mobility 2DEG. Also, we would like to thank \mbox{D. Wulferding} for discussions. This work was financially supported by the Cluster of Excellence QUEST. \mbox{J. I.} is supported by the MCYT (Spain) under grant MAT2011-24331 and ITN Grant 234970 (EU).
\end{acknowledgments}

\end{document}